\documentclass[pra,twocolumn,lengthcheck,showpacs]{revtex4}

\usepackage{amsmath}
\usepackage{amssymb}
\usepackage{amsthm}
\usepackage{graphicx}

\newcommand{\tr}{\operatorname{tr}}
\newcommand{\uinvnorm}{|\kern-2pt|\kern-2pt|}

\newcommand{\supp}{\operatorname{supp}}
\newcommand{\diam}{\operatorname{diam}}

\theoremstyle{plain}

\theoremstyle{definition}

\theoremstyle{remark}

\begin{document}

\title{Simulating adiabatic evolution of gapped spin systems}

\author{Tobias J.\ Osborne}
\email[]{Tobias.Osborne@rhul.ac.uk} \affiliation{Department of
Mathematics, Royal Holloway University of London, Egham, Surrey TW20
0EX, United Kingdom}


\date{\today}

\begin{abstract}
We show that adiabatic evolution of a low-dimensional lattice of
quantum spins with a spectral gap can be simulated efficiently. In
particular, we show that as long as the spectral gap $\Delta E$
between the ground state and the first excited state is any constant
independent of $n$, the total number of spins, then the ground-state
expectation values of local operators, such as correlation
functions, can be computed using polynomial space and time
resources. Our results also imply that the local ground-state
properties of any two spin models in the same quantum phase can be
efficiently obtained from each other. A consequence of these results
is that adiabatic quantum algorithms can be simulated efficiently if
the spectral gap doesn't scale with $n$. The simulation method we
describe takes place in the Heisenberg picture and does not make use
of the finitely correlated state/matrix product state formalism.
\end{abstract}

\maketitle

\section{Introduction}

The low-temperature physics of lattices of interacting quantum spins
is typically very complex. The computational cost of even
\emph{approximating} basic properties, such as the ground-state
energy eigenvalue, of these systems is prohibitive. Indeed, for 2D
lattices of interacting spins, the task of computing an
approximation to the ground-state energy eigenvalue correct to
within some polynomial confidence interval is fantastically
difficult --- this problem is complete for the complexity class {\sf
QMA}, which is the quantum version of the complexity class {\sf NP}
\cite{oliveira:2005a, kempe:2004a, kitaev:2002a}.

It might therefore seem that the computational task of approximating
the low-temperature behaviour of interacting quantum spins is
entirely hopeless. However, for physically realistic models, this is
not the case in practice. Many algorithms have been developed which
appear to provide efficient approximations to a wide variety of
local properties of physically realistic systems, such as
correlators, at low temperature. Perhaps the most successful of
these methods has been the family of algorithms based on the density
matrix renormalisation group (DMRG) (See \cite{schollwoeck:2005a}
and references therein for a review of the DMRG and description of
extensions.)

The DMRG is a remarkably flexible and adaptable algorithm, admitting
a slew of generalisations. Applications include: simulating dynamics
\cite{vidal:2003a, vidal:2003b}, dissipative systems
\cite{verstraete:2004b, zwolak:2004a}, disordered systems
\cite{paredes:2005a}, and higher dimensional lattices
\cite{verstraete:2004a}. At least part of the flexibility of the
DMRG is due to the fact that it is equivalent to a variational
minimisation over the space of \emph{finitely correlated states}
(FCS) \cite{fannes:1992a}. Hence, the methodology of the DMRG can be
adapted to any situation where the principle object of study, be it
an eigenstate or propagator, can be approximated using a FCS vector
on Hilbert space. An alternative to methods based on variations over
FCS has been recently proposed which appears to offer spectacular
computational speedups over the DMRG and relatives
\cite{vidal:2005a}.

In practice it appears that the DMRG and related algorithms can
efficiently obtain arbitrarily accurate approximations to the local
ground-state properties of a $1$D collection of interacting quantum
spins. However, at the current time, there is no satisfactory
understanding of the \emph{correctness} (i.e.\ will the DMRG
\emph{always} return a faithful approximation to the ground state
and not some other eigenstate) and the \emph{complexity} (i.e.,
assuming correctness, how much computational resources are required
to obtain a good approximation to a ground state) of the DMRG.

The correctness of the DMRG is far from obvious. This is because the
ground-state approximation obtained by the DMRG \emph{cannot} be
certified; the DMRG only returns an approximate ground-state
eigenvector and cannot guarantee that this vector is close to the
true ground state. It is therefore extremely desirable to determine
\emph{a priori} the class of systems for which the DMRG and
relatives provably return faithful approximations to the ground
state. The complexity \cite{endnote40} of the DMRG is also difficult
to ascertain. Assuming we could prove correctness of the DMRG for a
class of realistic physical systems, the actual complexity of the
DMRG depends subtly on many detailed properties of the system, such
as geometric entropy of the ground state, and nonconvexity of the
objective function which is minimised.

Recently this situation is changing \cite{hastings:2006a,
verstraete:2005a, osborne:2005d}. In \cite{hastings:2005b} an
analysis of the resource scaling of a DMRG-like algorithm to obtain
approximations to the ground states of $1$D gapped local models was
undertaken. This paper provides the first general subexponential
estimate for the time and space resource requirements of any
\emph{provably correct} method to compute approximations to the
ground states of gapped models; it was found that if the model is
gapped then resources scaling as $n^{c\log n}$, with $c$ some
constant, are sufficient to obtain and store a computational
representation of the ground state of a gapped local model
\cite{endnote41}. In \cite{verstraete:2005a} it was shown that the
ground state of some \cite{endnote42} \emph{critical} $1$D spin
models can be \emph{stored} efficiently. Unfortunately, there is
currently no theoretical argument which implies that these
approximations to the ground states can be obtained efficiently.
Indeed, the results of this paper imply that if such approximations
are obtained via adiabatic continuation then \emph{exponential}
computational resources may be required to obtain them. (Note,
however, that we can say nothing about the other methods to obtain
such FCS approximations.) Finally, in \cite{osborne:2005d} it was
shown that an approximation to the propagator for a $1$D lattice of
quantum spins can be obtained and stored (as a FCS vector) using
polynomial resources in $n$ and the error $\epsilon$ and exponential
resources in the time $|t|$. (It is straightforward to extend the
argument of \cite{osborne:2005d} to show an analogous result in
$2$D.)

There is at least one solid reason why we believe that DMRG-like
methods ought to provide a computationally efficient recipe to
compute approximations to the ground states of gapped systems.
Namely, we know that the ground-state correlation functions for any
gapped system are \emph{clustering} or rapidly decaying with
separation \cite{hastings:2004a, hastings:2004b,
nachtergaele:2005a}. This result, which is the natural analogue of
Fredenhagen's proof \cite{fredenhagen:1985a} of clustering for
relativistic quantum field theories, is especially impressive given
that it applies to an extremely wide class of quantum lattice
systems in low dimensions. As a consequence of clustering results we
conclude that gapped spin systems are essentially free --- an
intuition which is persuasively backed up by classical
renormalisation-group style argumentation  --- and thus can be
modelled as noninteracting \emph{effective} spins, which can be
simulated easily.

Another way of arriving at this conclusion is to think of
correlations as roughly ``measuring'' the degree of \emph{quantum}
correlations in the ground state. Since the amount of quantum
correlations in a quantum state limits the extent to which a state
can be approximated by a FCS \cite{verstraete:2005a}, we are
strongly encouraged to think that the clustering results may
actually imply that DMRG-like algorithms may converge rapidly for at
least some realistic gapped systems.

Unfortunately, knowing that the correlations decay is not enough
information to infer that the eigenstates are finitely correlated.
To understand this simply consider a \emph{generic quantum state}
\cite{hayden:2004b} which is a quantum state chosen uniformly from
Haar measure induced on state-space. A generic quantum state
exhibits rapidly decaying correlations (indeed, all $m$-point
correlation functions are essentially zero for $m < \frac{n}{2}$)
yet such a state is extremely entangled and cannot be efficiently
represented as a finitely correlated state. Nevertheless, it might
be argued that the results of \cite{hastings:2004a, hastings:2004b,
nachtergaele:2005a} avoid this counterexample because they prove
something stronger, namely \emph{exponential clustering}, which says
that the reduced density operator $\rho_{AB}$ of the ground state
for two arbitrarily large separated regions $A$ and $B$ is
indistinguishable from a product $\rho_A\otimes\rho_B$ when it is
used to compute expectations for \emph{product} observables $M_A
M_B$. Interestingly, a naive attempt to exploit this exponential
clustering runs into problems. The reason is that there exist highly
entangled states $\sigma_{AB}$, called \emph{data-hiding states},
which exhibit precisely these properties \cite{hayden:2004c}. Thus,
to prove that the ground state of a gapped local hamiltonian is
well-approximated by finitely correlated state with polynomial
resources we appear to need more information than that given by
correlation functions.

Despite some recent progress a solution to the fundamental problem,
namely, to prove correctness of \emph{any} algorithm which obtains
approximations to local ground-state properties for gapped $1$D
models and to further provide a \emph{polynomial} theoretical
worst-case estimate on the resource requirements such an algorithm,
still seems far away. Let us summarise the various approaches to
finding approximations to the ground state of a spin model and the
theoretical obstructions encountered in each of these approaches.

There are at least four ways to obtain an approximation to the
ground state of a quantum system: (i) variation over a class of
ansatz ground states; (ii) simulation of the thermalisation process
via imaginary time evolution or similar; (iii) approximation of the
convex set of reduced density operators of translation-invariant
quantum states; and (iv) adiabatic continuation from the ground
states of classical spin models. The DMRG is an example of the first
method, namely it is a variation over the class of FCS with fixed
auxiliary dimension. Unfortunately this variation is, in general,
nonconvex and it has been recently discovered \cite{eisert:2006a}
that hard instances for a closely related variation problem can be
constructed. Thus it seems likely that the DMRG is not correct in
general. The second approach, namely imaginary time evolution,
suffers from the shortcoming that an initial guess $|\Omega'\rangle$
for the ground state $|\Omega\rangle$ which has nontrivial overlap
with the actual ground state is required. If such an initial guess
is unavailable then the storage requirements of the imaginary time
evolution approach could be, in the worst case, exponential
\cite{endnote43}. It seems plausible that obtaining such a guess
could be as hard as solving the original problem. The third method
requires an exponentially good characterisation of the convex set of
reduced density operators of translation-invariant quantum states in
order to obtain $O(1)$ estimates for local operators. The final
method, which is the focus of this paper, suffers from the
limitation that it is not known if the ground state of an arbitrary
gapped spin model can be obtained via adiabatic continuation from a
classical model without encountering a quantum phase transition.
However, it has been recently proved \cite{yarotsky:2004a,
yarotsky:2005a, yarotsky:2005b} that in the \emph{neighbourhood} of
a classical spin model adiabatic continuation \emph{will} work.
Thus, using this approach, we are able to provide the first
polynomial estimates on the resource requirements of a correct
method to obtain a representation of the ground state of at least a
subclass of gapped models.

There is an intimate connection between simulating adiabatic
continuation for quantum lattice models and simulating quantum
computations \cite{aharonov:2004a, kempe:2004a}. Namely, if
adiabatic evolution for an arbitrary 2D lattice model with a gap
that scales as an inverse polynomial of the system size could be
simulated efficiently on a classical computer then \cite{endnote44}
{\sf BQP}$\subseteq${\sf P}, thus obviating the need to design and
engineer a quantum computer in the first place! Naturally, our
results are nowhere near strong enough to show the complexity class
inclusion {\sf BQP}$\subseteq${\sf P}, but they do have implications
for error correction methods for adiabatic quantum algorithms.

A complete theory of quantum error correction for adiabatic quantum
algorithms \cite{farhi:2001a} is still being developed. For example,
for general thermalisation decoherence, we really have no idea how
to calculate a fault-tolerance threshold for adiabatic quantum
algorithms (see \cite{nielsen:2000a} and \cite{preskillnotes} for a
discussion of quantum error correction and fault tolerance.).
Presumably a general quantum error-correcting code for a quantum
adiabatic algorithm would involve encoding the adiabatic evolution
in a larger system such that the minimum spectral gap encountered
along the evolution was larger \cite{endnote45}. This would mean it
would cost the environment more energy/unit time to induce a
transition from the ground state during the evolution (an
``error''). It is natural to assume that the gap could be boosted to
a large constant, independent of the number $n$ of spins, with a
polynomial increase in size. Our results show that if this were
possible then we could simulate adiabatic quantum algorithms
efficiently on a classical computer! Thus, conditioned on the strict
complexity class containment {\sf P} $\subset$ {\sf BQP}, we obtain
a bound on how large the gap could be boosted by encoding for
adiabatic quantum algorithms.

The method we develop in this paper is very closely related to the
method studied in \cite{wen:2005a}. In \cite{wen:2005a} the authors
investigate the evolution of local operators under a quasi-adiabatic
change in a local hamiltonian. As long as the hamiltonian has a
spectral gap throughout the evolution, it was found that local
operators remained local and thus it was possible to say that local
gauge invariance remains when two hamiltonians are in the same
phase. Our task is similar: we wish to understand the expectation
values of local operators in the ground state of a system that has
\emph{undergone} adiabatic evolution. We wish to show that the
computation of such expectation values can be done efficiently on a
classical computer as long as the smallest gap encountered during
the adiabatic evolution is $O(1)$. While this calculation can be
treated using quasi-adiabatic evolution and the methods developed in
\cite{wen:2005a} to study such evolutions, we prefer to study
\emph{exact} adiabatic evolution. We do this primarily in
anticipation of the application of these results to studying
entropy-area laws for systems in the same phase.

We provide an efficient computational method to compute the
expectation values of local operators in the ground states of
hamiltonians undergoing exact adiabatic evolution, a method which
works equally well for hamiltonians with spatially varying
interactions. Our method does not make use of the FCS formalism.
Rather, we develop our simulation method in the \emph{Heisenberg
picture}, where locality is manifest. Indeed, if we were to make use
of state representations in the Schr\"odinger picture, i.e.\ the 2D
FCS formalism (PEPS), we would be unable to apply our results
because even if we could construct PEPS approximations to the
adiabatically continued ground state it is currently unknown how to
efficiently extract expectation values of local operators from the
PEPS representation. We sidestep this issue by providing a
ground-state \emph{certificate} in the form of a specification of a
local hamiltonian which can be efficiently numerically simulated in
the Heisenberg picture to extract local expectation values.

The outline of this paper is as follows. We begin in
\S\ref{sec:form} by introducing the class of local hamiltonians we
consider and stating the problem we wish to solve. In
\S\ref{sec:efflocal} we then show how adiabatic evolution for
quantum lattices of spins can be described by unitary dynamics of an
\emph{effective} local hamiltonian. We use this effective dynamics
in \S\ref{sec:effsim} to construct an approximate local dynamics
which can then be used to efficiently extract local properties of
the adiabatically continued ground state. We conclude with some
discussion of our results in \S\ref{sec:disc}. We detail some simple
properties of compactly supported $C^\infty$ functions in
Appendix~\ref{app:cutoff}.

\section{Formulation}\label{sec:form}

In this section we introduce the Hilbert space and operator algebras
for the systems we consider. We define what we mean by strictly
local and approximately local hamiltonians. Finally, we specify the
computational task that will occupy us for the rest of this paper.

We consider quantum systems defined on a set of vertices $V$ with a
finite dimensional Hilbert space $\mathcal{H}_x$ attached to each
vertex $x\in V$. We always assume that $V$ is finite. (There are
some minor theoretical obstructions which currently preclude a
simple extension of our results to infinite lattices; we'll discuss
this in a further paper.) For $X\subset V$, the Hilbert space
associated to $X$ is the tensor product
$\mathcal{H}_X=\bigotimes_{x\in X}\mathcal{H}_x$, and the algebra of
observables on $X$ is denoted by
$\mathcal{A}_X=\mathcal{B}(\mathcal{H}_X)$, where
$\mathcal{B}(\mathcal{H}_X)$ denotes the $C^*$-algebra of bounded
operators on $\mathcal{H}_X$ with norm
\begin{equation}
\|A\| = \sup_{|\psi\rangle \in \mathcal{S}(\mathcal{H}_X)}
\|A|\psi\rangle\|,
\end{equation}
and $\mathcal{S}(\mathcal{H}_X)$ is the state space for
$\mathcal{H}_X$. We assume that $V$ is equipped with a metric $d$.
In the most common cases $V$ is the vertex set of a graph, and the
metric is given by the graph distance, $d(x,y)$, which is the length
of the shortest path of edges connecting $x$ and $y$ in the graph.
Finally, by tensoring with the unit operators on $Y\setminus X$, we
consider $\mathcal{A}_X$ as a subalgebra of $\mathcal{A}_Y$,
whenever $X\subset Y$.

We will, for the sake of clarity, introduce and describe our results
for a collection of $n$ distinguishable spin-$\frac{1}{2}$
particles. Thus, the Hilbert space $\mathcal{H}$ for our system is
given by $\mathcal{H} = \bigotimes_{j=0}^{n-1} \mathbb{C}^2$. We now
fix the metric for our vertex set $V$ to be that of a
low-dimensional periodic lattice $L$ of $n=m^\eta$ vertices, where
$m\in\mathbb{N}$ and $\eta$ is the dimension. Because the case
$\eta=2$ is the only really nontrivial case that interests us, we
fix $\eta=2$ from now on. We refer to vertices as \emph{sites} and
identify each site $v$ with its coordinates $\mathbf{j} = (j_x,
j_y)$. Because the lattice is periodic we identify coordinates:
$(j_x=m) \equiv (j_x = 0)$ and $(j_y=m) \equiv (j_y = 0)$. It is
entirely straightforward to generalise our results to
higher-dimensional lattices, higher dimensional spins, and to more
general lattices.

We consider a distinguished basis, the \emph{standard product
basis}, for $\mathcal{H}_{V}$ given by $|\mathbf{z}\rangle =
\bigotimes_{{j_x}=0}^{m-1}\bigotimes_{{j_y}=0}^{m-1}
|z_{(j_x,j_y)}\rangle$, $z_{\mathbf{j}}\in \mathbb{Z}/2\mathbb{Z}$.
We'll also have occasion to refer to a certain orthonormal basis for
$\mathcal{A}_{V}$: we denote by $ \sigma^{\boldsymbol{\alpha}} =
\bigotimes_{{j_x}=0}^{m-1}\bigotimes_{{j_y}=0}^{m-1}
\sigma_{(j_x,j_y)}^{\alpha_{(j_x,j_y)}}$, $\alpha_{\mathbf{j}}\in
\mathbb{Z}/4\mathbb{Z}$, the \emph{standard operator basis}, where $
\sigma^{0} = \left(\begin{smallmatrix} 1 & 0 \\
0 & 1
\end{smallmatrix}\right)$, $\sigma^{1} =  \left(\begin{smallmatrix} 0 & 1 \\ 1 & 0
\end{smallmatrix}\right)$, $\sigma^{2} = \left(\begin{smallmatrix} 0 & -i \\ i & 0
\end{smallmatrix}\right)$, and $\sigma^{3} = \left(\begin{smallmatrix} 1 & 0 \\ 0 & -1
\end{smallmatrix}\right)$, are the Pauli sigma matrices.

We define the \emph{support} $\supp(M)\subset V$ of an operator
$M\in\mathcal{A}_V$ to be the smallest subset $\Lambda \subset V$
such that $M\in\mathcal{A}_{\Lambda}$, i.e., the smallest subset
upon which $M$ acts nontrivially. Let $M\subset L$ and $N\subset L$.
We define the \emph{sumset} $M+N\subset L$ of $M$ and $N$ by $M+N =
\{\mathbf{x}+\mathbf{y}\,|\, \mathbf{x}\in M, \mathbf{y}\in N\}$
where the addition operation $\mathbf{x}+\mathbf{y}$ is inherited
from the standard addition on $L\equiv(\mathbb{Z}/m\mathbb{Z})\times
(\mathbb{Z}/m\mathbb{Z})$. This operation is the natural
generalisation of the convolution operation on the real numbers to
the finite group $L$. (It is fairly straightforward to generalise
these operations to more general graphs.)

We now introduce the family $H(s)$ of parameter-dependent
hamiltonians we are going to focus on. To define our family we'll
initially fix some parameter-dependent interaction term
$h(s)\in\mathcal{A}_V$ which has bounded norm \cite{endnote46}:
$\|h(s)\| \le O(1)$. We think of $h(s)$ as being ``centred'' on site
$\mathbf{0}$, i.e.\ we demand that $\mathbf{0}\in\supp(h(s))$. Our
family $H(s)$ of quantum systems is then defined by
\begin{equation}\label{eq:tihamdef}
H(s) = \sum_{\mathbf{j}\in L}
\mathcal{T}_y^{j_y}(\mathcal{T}_x^{j_x}(h(s))) = \sum_{\mathbf{j}\in
L} h_{\mathbf{j}}(s),
\end{equation}
where $\mathcal{T}_x$ (respectively, $\mathcal{T}_y$) is the unit
translation operator which translates the subsystems one site across
in the $x$ (respectively, $y$) direction, eg.,
\begin{equation}
\mathcal{T}_x\left(\bigotimes_{{j_x}=0}^{m-1}\bigotimes_{{j_y}=0}^{m-1}
\sigma_{(j_x,j_y)}^{\alpha_{(j_x,j_y)}}\right) =
\bigotimes_{{j_x}=0}^{m-1}\bigotimes_{{j_y}=0}^{m-1}
\sigma_{(j_x+1,j_y)}^{\alpha_{(j_x,j_y)}},
\end{equation}
and $h_{\mathbf{j}}(s) =
\mathcal{T}_y^{j_y}(\mathcal{T}_x^{j_x}(h(s)))$. While the
hamiltonian $H(s)$ generated by this construction is
translation-invariant, none of our subsequent calculations depend on
this fact in any serious way. Hence the results of this paper apply
equally to hamiltonians with spatially varying interactions.

We are going to make three simplifying assumptions about our
hamiltonian $H(s)$. The first is that $H(s)$ is assumed to be
\emph{strictly local} which means that $|\supp(h(s))|$ is an $O(1)$
constant. The second assumption we make is that the interaction
$h(s)$ that generates $H(s)$ can be written as $h(s) = h_0 + s h'$,
where $h_0$ and $h'$ are two operators with $O(1)$ norm. The final
assumption is that the ground state is unique and the spectral gap
$\Delta E(s)$ between the ground- and first-excited states for
$H(s)$ satisfies the inequality $\Delta E(s) \ge \Delta$, $\forall
s\in[0,1]$, where $\Delta E(s)$ is an $O(1)$ constant. Note that the
first two assumptions can be lifted with a little extra work,
however, the assumption that the gap $\Delta E(s)$ is an $O(1)$
constant cannot be relaxed: the simulation algorithm we present
scales exponentially with $\Delta E(s)$.

We will also have occasion to discuss \emph{approximately local}
hamiltonians. Such hamiltonians are obtained in the same way as in
(\ref{eq:tihamdef}), that is, we fix some initial interaction term
$k(s)$ which we then average over translates to generate our
hamiltonian $K(s)$. In this case, however, the initial interaction
term is allowed to have support equal to all of $V$. The only
constraint we make is that $k(s)$ must \emph{decay rapidly} which
means that $k(s)$ can be written as a sum:
\begin{equation}
k(s) = \sum_{\alpha=0}^{m-1} k_{\alpha}(s)
\end{equation}
where $\supp(k_{\alpha}(s)) = \Lambda_\alpha$, and $\Lambda_\alpha$
consists of all the sites within a distance $\alpha$ of site $0$,
i.e., $\Lambda_\alpha = \{\mathbf{j}\, | \, d(\mathbf{0},\mathbf{j})
\le \alpha \}$. As a result, $k_\alpha(s)$ is an operator with a
support (or ``radius'') consisting of $\alpha$ sites centred on site
$0$. The rapid decay condition is then that
\begin{equation}
\|k_{\alpha}(s)\| \le f(\alpha), \quad 0 \le \alpha < m.
\end{equation}
where $f(\alpha)$ is some rapidly decreasing function of $\alpha$.

We say that a hamiltonian $K(s)$ constructed from the interaction
$k(s)$ has \emph{rapid decay}. We write the final hamiltonian
resulting from this construction as
\begin{equation}\label{eq:ksdef}
K(s) = \sum_{\mathbf{j}\in L}\sum_{\alpha = 0}^{m-1}
k_{\mathbf{j},\alpha}(s),
\end{equation}
where $k_{\mathbf{j},\alpha}(s) =
\mathcal{T}_y^{j_y}(\mathcal{T}_x^{j_x}(k_{\alpha}(s)))$.

Finally, we set out the problem we aim to solve. We suppose $H(s)$
is a strictly local parameter-dependent hamiltonian for a $2$D
lattice of the form (\ref{eq:tihamdef}), with interaction $h(s)$
having $O(1)$ norm and $O(1)$ support. We assume that the ground
state $|\Omega(s)\rangle$ is unique and, further, that the spectral
gap $\Delta E(s)$ between the ground state and first excited state
satisfies $\Delta E(s) \ge \Delta$, $\forall s\in [0, 1]$, where
$\Delta$ is a constant independent of $n$. Finally, we suppose that
expectation values of arbitrary local operators $A\in\mathcal{A}_L$,
with $O(1)$ support, in the initial ground state $|\Omega(0)\rangle$
can be computed efficiently, i.e., $\omega_0(A) = \langle \Omega(0)|
A |\Omega(0)\rangle$ can be computed efficiently for all
$A\in\mathcal{A}_L$. This would be the case when, for example,
$H(0)$ is any regular classical hamiltonian, that is,
$[h_{\mathbf{j}}(s), h_{\mathbf{k}}(s)]=0$, $\forall \mathbf{j},
\mathbf{k} \in L$. Alternatively, this occurs when $H(s)$ has a
ground state which is exactly a $2$D finitely correlated state.
(When $H(s)$ has spatially varying interactions we must require that
the ground state of $H(0)$ is a \emph{known} product state. We need
to do this in order to avoid the constructions of
\cite{barahona:1982}, which show that computing the ground state of
a disordered classical systems is at least {\sf NP}-hard.)

Our approximation problem is therefore the following. First fix some
error $\epsilon$. Then our problem is to find an efficient
computational method to compute, for any local operator $A$ with
bounded support \cite{endnote47}, uniform approximants
$\omega'_{s}(A)$ to the exact expectation values $\omega_s(A) =
\langle \Omega(s)|A|\Omega(s)\rangle$. That is, our problem is to
efficiently compute $\omega'_{s}(A)$ so that $|\omega'_{s}(A) -
\omega_s(A)|<\epsilon$ for all $s\in[0,1]$ and for all bounded local
operators $A$ with bounded support.

The constraint that the observables whose expectation values are to
be simulated must have bounded support stems from the condition that
in the large-$n$ limit such operators should be elements of the
quasi-local algebra $\mathcal{A}_L$. We lose no generality in this
assumption when applying it to the simulation of quantum algorithms
because the answer that the algorithm computes should be encoded in
the ground state in such a way that it can be read out from the
expectation value of a local operator. It is also worth noting that
any correlation function involving a bounded number of subsystems
satisfies our definition of having bounded support.

Before we end this section we introduce some notation for
approximations. Because we have occasion to refer to functions for
which only \emph{bounds} on growth, derivatives, etc.\ are known it
is convenient to adopt the following notation. If we have two
quantities $A$ and $B$ then we use the notation $A\lesssim B$ to
denote the estimate $A\le CB$ for some constant $C$ which only
depends on unimportant quantities. In almost all the cases we
consider the only important quantity is $n$, the total number of
spins. Thus, unless we indicate otherwise, $A\lesssim B$ means that
$A\le CB$ for some $C$ independent of $n$. Because we'll be
interested in the consequences of allowing the minimum gap $\Delta$
to depend on $n$ we'll explicitly retain any dependence on $\Delta$
in our calculations.

\section{Effective local dynamics for exact adiabatic
evolution}\label{sec:efflocal}

In this section we study exact adiabatic evolution for quantum spin
systems. We show that if there is a gap throughout the evolution
then the exact adiabatic evolution is equivalent to unitary dynamics
generated by an approximately local hamiltonian.

We consider adiabatic quantum evolution generated by $H(s)$ as $s$
is varied adiabatically from $s=0$ to $s=1$. Thus we would like to
understand the ground state $|\Omega(s)\rangle$ of $H(s)$. We do
this by setting up a differential equation for $|\Omega(s)\rangle$:
\begin{equation}\label{eq:pformula}
\frac{d}{ds} |\Omega(s)\rangle = P'(s)|\Omega(s)\rangle,
\end{equation}
where $P'(s) = \frac{d}{ds}(|\Omega(s)\rangle\langle\Omega(s)|)$ and
we've set phases \cite{endnote48} so that $\langle
\Omega'(s)|\Omega(s)\rangle = 0$. Because $P'(s)$ is not
antihermitian the dynamics generated by this equation are not
unitary.

There are at least two ways to set up differential equations for
$|\Omega(s)\rangle$ which \emph{do} generate unitary dynamics. The
first is via \emph{exact adiabatic evolution} (see
\cite{avron:1987a, avron:1993a} for a rigourous discussion of rather
general results about exact adiabatic evolution):
\begin{equation}\label{eq:gpformula}
\frac{d}{ds} |\Omega(s)\rangle = -[P(s), P'(s)]|\Omega(s)\rangle.
\end{equation}
Because of the gap condition on $H(s)$, the ``hamiltonian''
$[P(s),P'(s)]$ for this dynamics is given by first-order stationary
perturbation theory:
\begin{multline}
[P(s), P'(s)] = |\Omega(s)\rangle \langle \Omega(s)| \frac{\partial
H(s)}{\partial s} \frac{\mathbb{I}}{\Omega(s)\mathbb{I}-H(s)} - \\
\frac{\mathbb{I}}{\Omega(s)\mathbb{I}-H(s)} \frac{\partial
H(s)}{\partial s} |\Omega(s)\rangle \langle \Omega(s)|,
\end{multline}
where $\Omega(s)$ is the ground-state energy of $H(s)$, and we
define $\frac{\mathbb{I}}{\Omega(s)\mathbb{I}-H(s)}$ via the
Moore-Penrose inverse:
$\frac{\mathbb{I}}{\Omega(s)\mathbb{I}-H(s)}|\Omega(s)\rangle =0$.

The other way, which we call \emph{effectively local exact adiabatic
evolution}, is obtained by rewriting $P'(s)$. We exploit the fact
that $H(s)$ has a spectral gap to find
\begin{equation}\label{eq:gsproj}
P(s) = \int_{-\infty}^{\infty} \chi_{\gamma}(t)
e^{-it\Omega(s)}e^{itH(s)}dt,
\end{equation}
where $\chi_{\gamma}(t)$ is an even real function whose fourier
transform $\widehat{\chi}_\gamma$ is $C^\infty$, has compact support
in $[-\gamma, \gamma]$, and is normalised so that
$\widehat{\chi}_\gamma(0) =1$. (See Appendix~\ref{app:cutoff} for a
description of $C^\infty$ cutoff functions and their properties.) We
must set $\gamma < \Delta$ to ensure that only the ground state
appears on the RHS of (\ref{eq:gsproj}). The formula
(\ref{eq:gsproj}) for $P(s)$ may be verified by writing $e^{itH}$ in
its eigenbasis and exploiting the $L_2$ unitarity of the fourier
transform.

We next use the Duhamel formula
\begin{equation*}
\frac{d}{ds}e^{itH(s)} =  i\int_0^t e^{iuH(s)}\frac{\partial
H(s)}{\partial s} e^{i(t-u)H(s)} du,
\end{equation*}
to rewrite (\ref{eq:pformula}):
\begin{widetext}
\begin{equation}
\frac{d}{ds} |\Omega(s)\rangle = -i \frac{d\Omega(s)}{ds}
\int_{-\infty}^{\infty} t\chi_\gamma(t) dt|\Omega(s)\rangle +
i\int_{-\infty}^{\infty}
\chi_{\gamma}(t)e^{-it\Omega(s)}\left(\int_0^t
\tau_u^{H(s)}\left(\frac{\partial H(s)}{\partial s}\right) du\right)
 e^{itH(s)}dt|\Omega(s)\rangle,
\end{equation}
\end{widetext}
where $\tau_u^{H(s)}(M) = e^{iuH(s)}Me^{-iuH(s)}$. Using the fact
that $\chi_\gamma(t)$ is an even function of $t$ and cancelling
phases we obtain
\begin{equation}
\frac{d}{ds} |\Omega(s)\rangle = i\int_{-\infty}^{\infty}
\chi_{\gamma}(t)\left(\int_0^t \tau_u^{H(s)}\left(\frac{\partial
H(s)}{\partial s}\right) du\right)
 dt|\Omega(s)\rangle.
\end{equation}
By integrating this expression for $\frac{d}{ds} |\Omega(s)\rangle$
in the energy eigenbasis of $H(s)$ and using the assumed gap
structure one can find that this expression is equivalent to the
usual expression obtained from first-order perturbation theory:
\begin{equation}
\frac{d}{ds} |\Omega(s)\rangle =
\frac{\mathbb{I}}{\Omega(s)\mathbb{I} - H(s)}\frac{\partial
H(s)}{\partial s} |\Omega(s)\rangle.
\end{equation}

Thanks to our assumed form of $H(s) = H_0 + s H'$, with $H' =
\sum_{\mathbf{j}\in L} h'_{\mathbf{j}} = \sum_{\mathbf{j}\in L}
\mathcal{T}_y^{j_y}(\mathcal{T}_x^{j_x}(h'))$, we notice that
$\frac{\partial H(s)}{\partial s} = \sum_{\mathbf{j}\in L}
h'_{\mathbf{j}}$, and we write
\begin{equation}\label{eq:adqevol}
\frac{d}{ds} |\Omega(s)\rangle = i\sum_{\mathbf{j}\in
L}\mathcal{F}_s(h_{\mathbf{j}}') |\Omega(s)\rangle,
\end{equation}
with initial condition that $|\Omega(0)\rangle$ is the ground state
of $H(0)$ and where $\mathcal{F}_s(M) = \int_{-\infty}^{\infty}
\chi_{\gamma}(t) \left(\int_0^t \tau_u^{H(s)}(M)\, du \right) dt$.

The equation (\ref{eq:adqevol}) tells us that $|\Omega(s)\rangle$
can be obtained from $|\Omega(0)\rangle$ by \emph{unitary} dynamics
according to the time-dependent hermitian hamiltonian $K(s) =
\sum_{\mathbf{j}\in L}\mathcal{F}_s(h_{\mathbf{j}}') =
\sum_{\mathbf{j}\in L} k_\mathbf{j}(s)$, where we write
$k_\mathbf{j}(s)=\mathcal{F}_s(h_{\mathbf{j}}')$. We also write
$k(s) = \mathcal{F}_s(h')$ for the interaction term $k(s)$ which
generates $K(s)$. Furthermore, we claim that $K(s)$ is
\emph{approximately local} for all $s\in[0,1]$.

The way to see that $K(s)$ is approximately local is to use the
standard Lieb-Robinson bound \cite{lieb:1972a, hastings:2004a,
nachtergaele:2005a, hastings:2005b}. The Lieb-Robinson bound reads
\begin{equation}\label{eq:liebrobinson}
\|[\tau_{t}^{H(s)}(A), B]\| \le |Y| e^{-v d(x,Y)}(e^{\kappa |t|}-1),
\end{equation}
for any two norm-$1$ operators $A\in\mathcal{A}_x$ and $B\in
\mathcal{A}_{Y}$, with $\{x\}\cap Y = \emptyset$ which are initially
separated by a distance $d(x,Y)$. The constants $v$ and $\kappa$ are
independent of $n$ and depend only on $\|h(s)\|$, which is an $O(1)$
constant.

What we do is define
\begin{equation}
k_{0}(s) = \mathcal{F}_s^{H_{\Lambda_0}(s)}(h')
\end{equation}
and
\begin{equation}
k_{\alpha}(s) = \mathcal{F}_s^{H_{\Lambda_\alpha}(s)}(h') -
\mathcal{F}_s^{H_{\Lambda_{\alpha-1}}(s)}(h'), \quad 0 < \alpha < m,
\end{equation}
where we define
\begin{equation}
\mathcal{F}_s^{H_{\Lambda_\alpha}(s)}(M) = \int_{-\infty}^{\infty}
\chi_{\gamma}(t) \left(\int_0^t \tau_u^{H_{\Lambda_\alpha}(s)}(M)
\,du \right)dt,
\end{equation}
with
\begin{equation}
H_{\Lambda_\alpha}(s) = \sum_{\mathbf{j}\in\Lambda_{\alpha}}
h_{\mathbf{j}}(s),
\end{equation}
where $\Lambda_\alpha = \{\mathbf{j}\, | \, d(\mathbf{0},\mathbf{j})
\le \alpha \}$. Obviously $k_{\alpha}(s)$ has support
$\supp(k_{\alpha}(s)) = \Lambda_\alpha+\supp(h')$. Also note that
$k(s) = \sum_{\alpha=0}^{m-1} k_{\alpha}(s)$ (recall that $m$ is the
diameter of the lattice).

We now show how the Lieb-Robinson bound provides an estimate on the
decay of $\|k_\alpha(s)\|$. Firstly, we rewrite the Lieb-Robinson
bound (\ref{eq:liebrobinson}) so that it is more useful:
\begin{widetext}
\begin{equation}\label{eq:liebrobinson2}
\begin{split}
\|\tau_t^{H_{\Lambda_\alpha}}(A) -
\tau_t^{H_{\Lambda_{\alpha-1}}}(A)\| &= \left\|\int_{0}^t ds\,
\frac{d}{d{t'}} (\tau_{t'}^{H_{\Lambda_{\alpha-1}}}(\tau_{t-t'}^{H_{\Lambda_{\alpha}}}(A))) \right\| \\
&= \left\| \int_{0}^t d{t'}\,
\tau_{t'}^{H_{\Lambda_{\alpha-1}}}([H_{\Lambda_\alpha}-H_{\Lambda_{\alpha-1}}, \tau_{t-{t'}}^{H_{\Lambda_\alpha}}(A)]) \right\| \\
&\le \int_0^{|t|} d{t'}\,
\|[H_{\Lambda_\alpha}-H_{\Lambda_{\alpha-1}},
\tau_{{t'}}^{H_{\Lambda_\alpha}}(A)]\| \\
&\le 2\int_0^{|t|} d{t'}\,
\|[H_{\Lambda_\alpha}-H_{\Lambda_{\alpha-1}}]\| e^{-v
\alpha+\kappa|t'|} \\
&\lesssim \alpha e^{\kappa|t| - v\alpha},
\end{split}
\end{equation}
\end{widetext}
where we used the fundamental theorem of calculus to get the first
line, the triangle inequality and unitary invariance of the norm to
get the third line, we substituted the Lieb-Robinson bound
(\ref{eq:liebrobinson}) in fourth line, and we integrated the bound
to get the fourth line. The $\alpha$ term in the fourth line comes
from the fact that the operator
$H_{\Lambda_\alpha}-H_{\Lambda_{\alpha-1}}$ consists of $\alpha$
terms (the number of terms crossing the boundary). The Lieb-Robinson
bound, in this form, says that the evolution of $A$ with respect to
$H_{\Lambda_\alpha}$ is almost the same as that for
$H_{\Lambda_{\alpha-1}}$, i.e., the boundary effects are unimportant
for short times.

\begin{figure*}
\begin{center}
\includegraphics{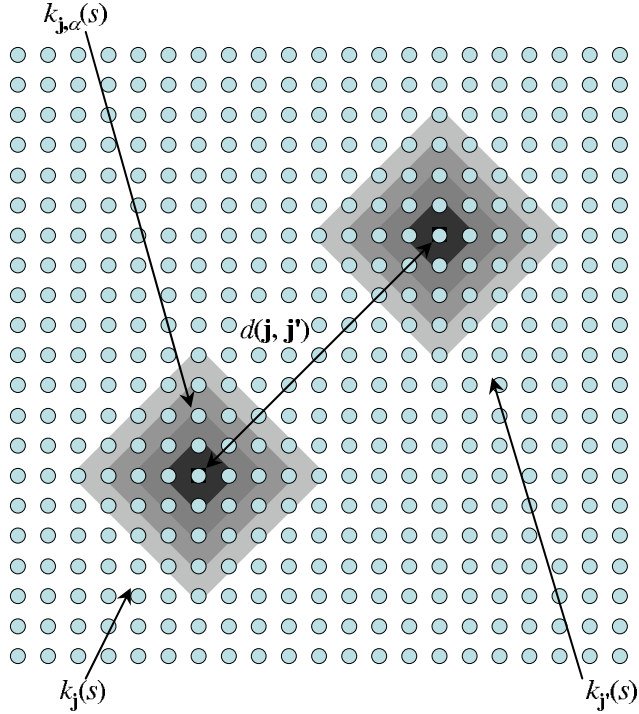}
\caption{Illustration of the rapidly decaying interactions for the
effectively local hamiltonian for exact adiabatic
evolution.}\label{fig:hint}
\end{center}
\end{figure*}

Now consider
\begin{widetext}
\begin{equation}\label{eq:kbound1}
\begin{split}
\|k_\alpha(s)\| &= \left\|\int_{-\infty}^{\infty} \chi_{\gamma}(t)
\left(\int_0^t \left(\tau_u^{H_{\Lambda_\alpha}(s)}(h') -
\tau_u^{H_{\Lambda_{\alpha-1}}(s)}(h') \right)du \right)dt\right\| \\
&\le 2\int_{0}^{\infty} |\chi_{\gamma}(t)| \left(\int_0^t
\left\|\tau_u^{H_{\Lambda_\alpha}(s)}(h') -
\tau_u^{H_{\Lambda_{\alpha-1}}(s)}(h') \right\|du\right) dt \\
&\le 2\int_{0}^{\infty} |\chi_{\gamma}(t)| \left(\int_0^t
\min\{2\|h'\|, c\alpha e^{\kappa |u| - v\alpha}\}du\right) dt \\
&\lesssim \alpha\int_{0}^{c\alpha}|\chi_\gamma(t)|e^{\kappa |t| -
v\alpha}dt + \int_{c\alpha}^\infty |\chi_\gamma(t)||t|dt \\
&\lesssim \alpha\int_{0}^{c\alpha} e^{\kappa t - v\alpha} dt +
\int_{c\alpha}^\infty \frac{1}{\gamma^l |t|^{l-1}}dt, \quad \forall l > 1 \\
&\lesssim \alpha e^{(\kappa c - v)\alpha} +
\frac{1}{c\gamma^l\alpha^{l-1}}, \quad \forall l > 1,
\end{split}
\end{equation}
\end{widetext}
where to get the second line we applied the triangle inequality, to
get the third line we applied the Lieb-Robinson bound in the form
(\ref{eq:liebrobinson2}) with $\alpha =
\diam(\Lambda_{\alpha})+\text{const.}$ (we've dropped the dependance
of the interactions $H_{\Lambda_\alpha}(s)$ on the parameter $s$
because for these inequalities the evolution is independent of the
parameter $s$), in the third line we've broken the integral into two
pieces and applied the different regimes of the Lieb-Robinson bound
separately with $c$ some constant \cite{endnote49} to be chosen
later, and in the final line we applied the decay estimates on
$\chi_\gamma(t)$ (see Appendix~\ref{app:cutoff} for a derivation of
these estimates). Thus, by choosing $c < v/\kappa$ we see that
$\|k_\alpha(s)\|$ is decaying faster than the inverse of any
polynomial in $\alpha$ for $\alpha \gtrsim 1/\gamma$, i.e., for
$\alpha
> c/\Delta$, where $c$ is some constant. In this way we see that
exact adiabatic evolution can be thought of as unitary dynamics
according to the paramater-dependent hamiltonian $K(s)$ which is
approximately local with respect to the metric $d$ on the lattice.
For an illustration of the interactions of $K(s)$ see
Figure~\ref{fig:hint}.

\section{Efficient simulation of adiabatic evolution}\label{sec:effsim}

In this section we apply a Lieb-Robinson bound to show that dynamics
according to effectively local exact adiabatic evolution keep local
operators approximately local and hence show that expectation values
of local operators in adiabatically evolved ground states can be
computed efficiently.

Recall that we can write the ground state $|\Omega(s)\rangle$ by
integrating (\ref{eq:adqevol}) as
\begin{equation}
|\Omega(s)\rangle = \mathcal{U}(s;0)|\Omega(0)\rangle,
\end{equation}
where
\begin{equation}
\mathcal{U}(s;0) = \mathcal{T}e^{i\int^s_0 K(s') ds'},
\end{equation}
and $\mathcal{T}$ denotes the time-ordering operation. Our objective
is to uniformly approximate
\begin{equation}
\omega_s(A) = \langle \Omega(0)|\mathcal{U}^\dag(s;0) A
\mathcal{U}(s;0)|\Omega(0)\rangle,
\end{equation}
for all $s\in[0,1]$. The way we do this is to show that the operator
$A(s) \equiv \mathcal{U}^\dag(s;0) A \mathcal{U}(s;0)$ remains
approximately local for all $s\in[0,1]$ and use the assumed fact
that $\omega_0(B)$ can be computed efficiently for all local
operators $B$. For simplicity we assume that the operator $A$ is
located at the origin and has support $|\supp(A)| = 1$. It is easy
to extend the results of this section to apply to operators with
bounded support on disconnected regions, such as correlators.

We now study the locality of $A(s)$. What we do is first show that
$A(s)$ can be uniformly approximated in operator norm by the series
of approximants
\begin{equation}
A_{\alpha}(s) \equiv \mathcal{V}^\dag_{\Lambda_\alpha}(s;0) A
\mathcal{V}_{\Lambda_\alpha}(s;0),
\end{equation}
where $\mathcal{V}_{\Lambda_\alpha}(s;0)$ satisfies the differential
equation
\begin{equation}
\frac{d}{ds}\mathcal{V}_{\Lambda_\alpha}(s;0) = i\sum_{\mathbf{j}\in
\Lambda_\alpha}\mathcal{F}_s(h_{\mathbf{j}}')\mathcal{V}_{\Lambda_\alpha}(s;0)
= iK_{\Lambda_\alpha}(s)\mathcal{V}_{\Lambda_\alpha}(s;0),
\end{equation}
with $\mathcal{V}_{\Lambda_\alpha}(0;0) = \mathbb{I}$ and where
$K_{\Lambda_\alpha}(s) = \sum_{\mathbf{j}\in
\Lambda_\alpha}\mathcal{F}_s(h_{\mathbf{j}}')$ and $\Lambda_\alpha =
\{\mathbf{j}\, | \, d(\mathbf{0},\mathbf{j}) \le \alpha \}$. In
words: the approximation $A_{\alpha}(s)$ is that operator obtained
by evolving $A$ with respect only to those interaction terms in
$K(s)$ whose centres are within a distance $\alpha$ of $A$.
Naturally this means that $A_{m-1}(s)=A(s)$. We use a Lieb-Robinson
bound to show that $\|A(s)-A_{\alpha}(s)\|$ is rapidly decaying.

To show this we prove $\|\tau_{s;0}^{K(s)}(A) -
\tau_{s;0}^{K_{\Lambda_\alpha}(s)}(A)\|$ is small for $|s| \le 1$
and large constant $\alpha$ where $\tau_{s;s'}^{K(s)}(M) =
\mathcal{U}^\dag(s;s')M\mathcal{U}(s;s')$. To make this expression
easier to deal with, and to more explicitly relate it to
group-velocity bounds, we rewrite it:
\begin{widetext}
\begin{equation}\label{eq:hamapprox}
\begin{split}
\|\tau_{s;0}^{K(s)}(A) - \tau_{s;0}^{K_{\Lambda_\alpha}(s)}(A)\| &=
\left\|\int_{0}^s ds'\,
\frac{d}{ds'} (\tau_{s';0}^{K_{\Lambda_\alpha}(s')}(\tau_{s;s'}^{K(s)}(A))) \right\| \\
&= \left\| \int_{0}^s ds'\,
\tau_{s';0}^{K_{\Lambda_\alpha}(s')}([K_{\Lambda_\alpha^c}(s'), \tau_{s;s'}^{K(s)}(A)]) \right\| \\
&\le \int_0^{|s|} ds'\, \|[K_{{\Lambda_\alpha}^c}(s'),
\tau_{s;s'}^{K(s)}(A)]\|,
\end{split}
\end{equation}
\end{widetext}
where $K_{{\Lambda_\alpha}^c}(s) = \sum_{\mathbf{j}\in L\setminus
{\Lambda_\alpha}} k_{\mathbf{j}}(s)$.

We now apply a general Lieb-Robinson bound recently proved in
\cite{hastings:2005b}. In order to apply the Lieb-Robinson bound of
\cite{hastings:2005b} we need to establish that our hamiltonian
$K(s)$ satisfies the conditions of Assumption 2.2 of
\cite{hastings:2005b}, which in our case reads
\begin{equation}
\sum_{\alpha=0}^{m-1} \|k_\alpha\|
(1+2\alpha(\alpha+1))^2(2+2\alpha)^\eta \le s_1,
\end{equation}
where $\eta$ is a positive constant and $s_1$ is some constant. We
need to ensure that the sum on the left evaluates to a constant
instead of diverging. The only flexibility we have is to choose a
decay estimate for $\|k_\alpha\|$ which is strong enough to
overwhelm the polynomial in $\alpha$ it is multiplied by. The
highest power of $\alpha$ appearing in this sum is
$\alpha^{4+\eta}$. Thus we use the decay estimate (\ref{eq:kbound1})
and choose $l\ge 7+\eta$, and so we find that this constant $s_1$
equates to
\begin{equation}
s_1 = \sigma(l)/\gamma^{l},
\end{equation}
where
\begin{equation}
\sigma(l) = c_l\sum_{\alpha=1}^{m-1}
\frac{(1+2\alpha(\alpha+1))^2(2+2\alpha)^\eta}{\alpha^{l-1}},
\end{equation}
and $c_l$ is the constant arising from the estimate
(\ref{eq:kbound1}), and $l$ is any chosen power $l\ge7+\eta$. (The
proof of the general Lieb-Robinson bound described in
\cite{hastings:2005b} is easily extended to cover
parameter-dependent hamiltonians such as $K(s)$.) This reads
\begin{equation}
\|[\tau_{s}^{K(s)}(A), B]\| \le
\frac{\rho(l)|Y|\left(e^{\frac{\sigma(l)}{\gamma^{l}}|s|}-1\right)}{(1+d(x,Y))^l},
\quad \forall l>7+\eta,
\end{equation}
for any two norm-$1$ operators $A\in\mathcal{A}_x$ and $B\in
\mathcal{A}_{Y}$, with $\{x\}\cap Y = \emptyset$ which are initially
separated by a distance $d(x,Y)$ and $\rho(l)$ is a constant which
depends only on $l$. The constant $v$ is independent of $n$ and
depends only on $\|h(s)\|$. We isolate the dependence of this bound
on the minimum gap $\gamma$ by defining
\begin{equation}
g(\gamma, l) =
\rho(l)\left(e^{\frac{\sigma(l)}{\gamma^{l}}|s|}-1\right).
\end{equation}
Note that we are going to systematically redefine this function in
our subsequent derivations to absorb extra constants and occurrences
of $\gamma$. With this bound we first obtain an upper bound on
$\|[\tau_{s,s'}^{K(s)}(A), k_{\mathbf{j},\alpha}(s')]\|$ (recall
that the operators $k_{\mathbf{j},\alpha}(s)$ are defined via
Eq.~(\ref{eq:ksdef})):
\begin{equation}
\|[A(s), k_{\mathbf{j},\alpha}(s)]\| \le
\begin{cases} \frac{g(\gamma,l)(1+2\alpha(\alpha+1))\|k_{\alpha}(s)\|}{(1+\delta-\alpha)^l}, \quad \alpha < \delta \\ 2\|A\| \|k_{\alpha}(s)\|, \quad \alpha \ge \delta, \end{cases}
\end{equation}
where $\delta = d(\mathbf{0}, \mathbf{j})$ and
$(1+2\alpha(\alpha+1))= |\Lambda_\alpha(\mathbf{j})| = |\{
\mathbf{x}\,|\, d(\mathbf{j}, \mathbf{x}) \le \alpha\}|$. We use the
estimate (\ref{eq:kbound1}) and redefine $g(\gamma,l)$ to find the
upper bound
\begin{equation}
\|[A(s), k_{\mathbf{j},\alpha}(s)]\| \le
\begin{cases} \frac{g(\gamma,l)(1+2\alpha(\alpha+1))}{\alpha^{l+2}(1+\delta-\alpha)^l}, \quad \alpha < \delta \\ 2\|A\| \|k_{\alpha}(s)\|, \quad \alpha \ge \delta. \end{cases}
\end{equation}
We next find the minimum of the denominator
$\alpha^{l+2}(1+\delta-\alpha)^l$ on the interval $1\le\alpha\le
\delta$, which is $\delta^l$, and redefine $g(\gamma,l)$ to arrive
at the final upper bound
\begin{equation}\label{eq:kjabound}
\|[A(s), k_{\mathbf{j},\alpha}(s)]\| \le
\begin{cases} \frac{g(\gamma,l)}{\delta^l}, \quad \alpha < \delta \\ \frac{2c_l\|A\|}{\gamma^{l+1}\alpha^l}, \quad \alpha \ge \delta. \end{cases}
\end{equation}

Thus, by choosing the centre $\mathbf{j}$ far enough away from the
centre $\mathbf{0}$ of $A(s)$ we find the behaviour
\begin{equation}
\|[A(s), k_{\mathbf{j},\alpha}(s)]\| \lesssim \frac{g(\gamma,l)}{
d(\mathbf{0},\mathbf{j})^l}, \quad \forall l>1,
\end{equation}
i.e., the quantity $\|[A(s), k_{\mathbf{j},\alpha}(s)]\|$ decays
faster than any polynomial in $d(\mathbf{0},\mathbf{j})$.

We next use our upper bound (\ref{eq:kjabound}) to obtain an upper
bound on $\|[A(s), k_{\mathbf{j}}(s)]\|$:
\begin{equation}\label{eq:1stadecay}
\begin{split}
\|[A(s), k_{\mathbf{j}}(s)]\| &\le \sum_{\alpha=0}^{m-1} \|[A(s),
k_{\mathbf{j},\alpha}(s)]\| \\
&\le \frac{g(\gamma,l)}{\delta^{l-1}} + \sum_{\alpha=\delta}^{m-1}
\frac{2c_{l+1}\|A\|}{\gamma^{l+1}\alpha^l} \\
&\le \frac{g(\gamma,l)}{\delta^{l-1}},
\end{split}
\end{equation}
where we've redefined $g(\gamma,l)$ in the last line.

Now we use the decay estimate (\ref{eq:1stadecay}) in
(\ref{eq:hamapprox}) to provide an upper bound for
$\|A(s)-A_{\alpha}(s)\|$:
\begin{equation}\label{eq:adecay1}
\begin{split}
\|A(s)-A_{\alpha}(s)\| &\le \sum_{\mathbf{j} \in L\setminus
\Lambda_\alpha}\int_{0}^{1} ds \|[A(s), k_{\mathbf{j}}(s)]\| \\
&\le \sum_{\delta = \alpha}^{m-1} \frac{(1+2\delta(\delta+1))
g(\gamma,l)}{\delta^{l-1}} \\
&\le \frac{g(\gamma,l)}{\alpha^{l-4}},
\end{split}
\end{equation}
where we've redefined $g(\gamma,l)$.

So, as long as $\alpha$ is chosen to be so large that it overwhelms
the $O(1)$ constant $g(\gamma,l)$ we find that
$\|A(s)-A_{\alpha}(s)\|$ can be made to decay faster than any
polynomial in $\alpha$, and hence, can be made as small as desired.
Thus there exists some constant $\alpha$ such that
$\|A(s)-A_{\alpha}(s)\| < \epsilon$. Note that, because $k(s)$ has
support throughout $L$, $A_\alpha(s)$ has support throughout $L$.

In order to provide a simulation method to compute approximations to
ground-state expectation values $\omega_s(A)$ we need to show that
$A_\alpha(s)$ can be approximated by an operator with support only
on a constant number of sites around $\supp(A)=\mathbf{0}$. The way
we do this is to show that $A_\alpha(s)$ is operator-norm close to
\begin{equation}
\widetilde{A}_{\alpha,\beta}(s) =
\widetilde{\mathcal{V}}^\dag_{\Lambda_{\alpha,\beta}}(s) A
\widetilde{\mathcal{V}}_{\Lambda_{\alpha,\beta}}(s),
\end{equation}
where $\widetilde{\mathcal{V}}_{\Lambda_{\alpha,\beta}}$ satisfies
the differential equation
\begin{multline}\label{eq:alphabetav}
\frac{d}{ds}\widetilde{\mathcal{V}}_{\Lambda_{\alpha,\beta}}(s) = \\
i\sum_{\mathbf{j}\in
\Lambda_{\alpha}}\widetilde{\mathcal{F}}_{\mathbf{j},s}^{H_{\Lambda_\beta(\mathbf{j})}}(h_{\mathbf{j}}')\widetilde{\mathcal{V}}_{\Lambda_{\alpha,\beta}}(s)
=
i\widetilde{K}_{\Lambda_{\alpha,\beta}}(s)\widetilde{\mathcal{V}}_{\Lambda_{\alpha,\beta}}(s),
\end{multline}
and
\begin{equation}
\widetilde{\mathcal{F}}_{\mathbf{j},s}^{H_{\Lambda_\beta(\mathbf{j})}}(h_{\mathbf{j}}')
= \int_{-\infty}^{\infty} \chi_{\gamma}(t) \left(\int_0^t
\tau_u^{H_{\Lambda_\beta(\mathbf{j})}(s)}(h_{\mathbf{j}}') du
\right)dt,
\end{equation}
with $\Lambda_\beta(\mathbf{j}) = \{ \mathbf{x}\,|\, d(\mathbf{j},
\mathbf{x}) \le \beta\}$.

To show that $\widetilde{A}_{\alpha,\beta}(s)$ is close to
$A_{\alpha}(s)$ we first exploit the general inequality
\begin{equation}
\|\mathcal{V}_{\Lambda_\alpha}(s)-\widetilde{\mathcal{V}}_{\Lambda_{\alpha,\beta}}(s)\|
\le \int_{0}^{|s|}
\|K_{{\Lambda_\alpha}}(s')-\widetilde{K}_{\Lambda_{\alpha,\beta}}(s')\|
ds'
\end{equation}
which is proved, for example, by exploiting the Lie-Trotter
expansion, and then upper-bound the right-hand side using the
triangle inequality by
\begin{multline}
\int_{0}^{|s|}
\|K_{{\Lambda_\alpha}}(s')-\widetilde{K}_{\Lambda_{\alpha,\beta}}(s')\|
ds' \le \\ \sum_{\mathbf{j}\in \Lambda_\alpha} \int _{0}^{|s|}
\|k_{\mathbf{j}}(s')-\widetilde{k}_{\mathbf{j},\beta}(s')\| ds',
\end{multline}
where $\widetilde{k}_{\mathbf{j},\beta}(s) =
\widetilde{\mathcal{F}}_{\mathbf{j},s}^{H_{\Lambda_\beta(\mathbf{j})}}(h_{\mathbf{j}}')$.
We can upper-bound the integral on the right-hand side by using an
argument identical to the one used to show (\ref{eq:kbound1}). We
thus obtain
\begin{equation}\label{eq:aalphadecay}
\begin{split}
\sum_{\mathbf{j}\in \Lambda_\alpha} \int _{0}^{|s|}
\|k_{\mathbf{j}}(s')-\widetilde{k}_{\mathbf{j}, \beta}(s')\| ds'
&\lesssim \sum_{\mathbf{j}\in \Lambda_\alpha}
\frac{1}{\gamma^l\beta^{l-1}} \\
&\lesssim \frac{\alpha^2}{\gamma^l\beta^{l-1}},
\end{split}
\end{equation}
where $l$ is any power, and we've used the fact that the number of
sites in $\Lambda_\alpha$ is given by $1+2\alpha(\alpha+1)$. By
choosing $\beta \gtrsim \alpha$ we find that
$\mathcal{V}_{\Lambda_\alpha}(s)$ can be made as close as desired to
$\widetilde{\mathcal{V}}_{\Lambda_{\alpha, \beta}}(s)$.

To obtain closeness of our final approximation
$\widetilde{A}_{\alpha,\beta}(s)$ to $A(s)$ we use the triangle
inequality
\begin{equation}
\begin{split}
\|A(s)-\widetilde{A}_{\alpha,\beta}(s)\| &\le \|A(s)-A_\alpha(s)\| +
\|A_\alpha(s)-\widetilde{A}_{\alpha,\beta}(s)\| \\
&\le \frac{g(\gamma, l)}{\alpha^{l}} +
\frac{\alpha^2}{\gamma^{l'}\beta^{l'-1}},
\end{split}
\end{equation}
where we've used the upper bound (\ref{eq:adecay1}) with an adjusted
value of $l$ and we've also used (\ref{eq:aalphadecay}) with an
appropriate choice of power $l'$. We therefore find that it is
sufficient, for a given constant $\epsilon$ to choose large (but
$O(1)$) $\alpha$ and $\beta$ so that
\begin{equation}\label{eq:finalaestimate}
\|A(s)-\widetilde{A}_\alpha(s)\| \le \epsilon.
\end{equation}

The actual values of $\alpha$ and $\beta$ required to reduce the
error (\ref{eq:finalaestimate}) to below $\epsilon$ scales better
than linearly with $w=\max(g(\gamma,l), 1/\gamma)$, where $\gamma$
is a constant multiplied by the minimum energy $\Delta E$
encountered along the adiabatic path. Thus the support of the final
approximation $\widetilde{A}_{\alpha,\beta}(s)$ is given, in the
worst case, by $\supp(\widetilde{A}_{\alpha,\beta}(s)) \lesssim w$.
Note that $w$ depends, via $g(\gamma,l)$, exponentially on
$1/\gamma$, i.e., the inverse energy gap.

Because the final approximation $\widetilde{A}_{\alpha,\beta}(s)$
can be computed via integrating (\ref{eq:alphabetav}), and by
noticing that this integration can be performed by restricting our
attention to the finite-dimensional subalgebra $\mathcal{A}_{W}$,
where $W=\supp(\widetilde{A}_{\alpha,\beta}(s))$, we see that
$\widetilde{A}_{\alpha,\beta}(s)$ can be computed using resources
which scale as $2^{cw}$, with $c$ some constant.

\section{Discussion}\label{sec:disc}

In this paper we have shown how to efficiently calculate the
ground-state expectation values of local operators with constant
support for gapped adiabatically evolving spin systems. In order to
provide our simulation method we reduced the problem to showing that
under exact adiabatic evolution the expectation value of a local
operator can be computed from the expectation value of an
approximately local operator in the unevolved ground state. Given
this observation we then argued that if it is easy to compute
expectation values of local operators in the original ground state
then one could approximate the desired expectation values
arbitrarily well by using time and space resources that scale with
the inverse gap.

Our approach has several shortcomings. The first is that the scaling
of the simulation resources with the error $\epsilon$ scales faster
than $2^{1/\epsilon}$. This means that if the expectation value of
an operator which is a sum of many local operators is desired then
our simulation method may require superpolynomial resources. For
example, if the expectation value of the \emph{total} magnetisation
$M = \sum_{\mathbf{j}\in L} \sigma_j^z$ (as opposed to the more
traditional \emph{average} magnetisation $m = M/n$) is required to
some accuracy $\epsilon$ then our simulation method will require
superpolynomial resources. This is not entirely unexpected, after
all, in the thermodynamic limit such operators are unbounded and
cannot be approximated at all. Another manifestation of this
shortcoming is that if the expectation values of the local operators
are required to an accuracy which scales as $\epsilon < 1/n$ then
our method may require superpolynomial resources. These problems do
not manifest themselves for the applications we have in mind.
Namely, when applied to the calculation of average properties of two
states in the same quantum phase we only require accuracy to some
small constant $\epsilon$ which doesn't scale with the system size,
and when applied to simulating adiabatic quantum algorithms we only
need $\epsilon$ to scale as a constant in order to read out the
answer of the algorithm.

The second shortcoming of our method is that, by the current method,
we are unable to directly approximate the scaling of the geometric
entropy \cite{endnote50} $S_\Lambda$ with $\Lambda$. The reason for
this is that our current method approximates $\rho_\Lambda(s)$ by
calculating approximations to all the expectation values of a basis
of operators for $\mathcal{A}_\Lambda$. Because we are computing
\emph{approximations} to expectation values we end up computing only
an \emph{approximation} $\widetilde{\rho}_{\Lambda}(s)$ to
$\rho_\Lambda(s)$. The best continuity result available for the von
Neumann entropy is Fannes inequality (see, for example,
\cite{nielsen:2000a}) for a derivation) which implies that the error
in the approximation $\widetilde{S}_\Lambda$ calculated from
$\widetilde{\rho}_{\Lambda}(s)$ grows larger as $\Lambda$ increases.
We'll describe an approach to this problem using exact adiabatic
evolution in a future paper.

\begin{figure*}
\begin{center}
\includegraphics{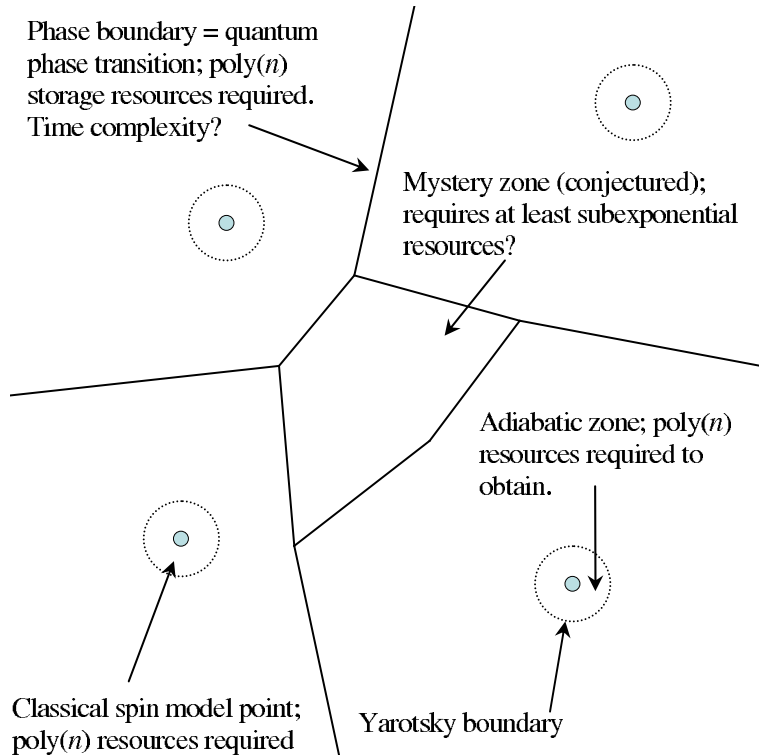}
\caption{Conjectured diagram of the space of local spin models and
the associated computational resources required to compute
approximations to local ground-state properties.}\label{fig:phases}
\end{center}
\end{figure*}

The principle characteristic of our approach is that approximations
are made in the \emph{Heisenberg picture}. What we mean here is that
instead of approximating the evolved quantum state of the spin
system in operator norm we instead compute approximations to the
evolved local operators. We should expect this strategy to be
successful because the locality of the interactions in the
hamiltonian doesn't manifest itself in the Schr{\"o}dinger picture
but, thanks to the Lieb-Robinson bound, it is precisely clear what
locality implies for local operators in the Heisenberg picture.
Because in the thermodynamic limit we are only able to physically
access local operators (such as average magnetisation and
correlators) this approach doesn't lead to any loss of generality
over computations carried out in the Schr{\"o}dinger picture.

It is possible that our analysis acutally applies to all gapped spin
models. This is because it is possible that any gapped spin model is
adiabatically connected \cite{endnote51} to a classical spin model
with trivial ground state. Classical renormalisation-group style
argumentation certainly seems to back this statement up: after all,
we know that the RG fixed points are either trivial (classical) or
quantum critical points. However, there is as yet no rigourous
general proof of this statement for quantum spin systems.

We would like to suggest that the following description of the space
of local (translation-invariant) spin models is correct. Firstly, in
this space there are many distinguished points, classical spin
systems, where the ground state can be calculated trivially. Around
each of these points is a small region in hamiltonian space of
hamiltonians which are provably adiabatically connected to the
classical spin model points \cite{yarotsky:2004a, yarotsky:2005a,
yarotsky:2005b}. In these regions we have shown that the local
ground-state properties can be determined efficiently. Outside these
small regions there are other regions which may or may not be
adiabatically connected to the classical spin model points where the
hamiltonians are gapped. In these regions it is known that the local
ground-state properties can be calculated using subexponential
resources \cite{hastings:2005b}. On the boundaries between the
quantum phases there are quantum critical walls. For these points,
in $1$D, it is known that an approximation to the ground state as a
finitely correlated state can be stored using polynomial space
\cite{verstraete:2005a}. It is not known if these approximations can
be obtained efficiently. This picture is summarised in
Figure~\ref{fig:phases}.

\subsection*{Acknowledgments}
I would like to thank Jens Eisert, Matthew Hastings, Jiannis Pachos,
Tony Short, Barbara Terhal, David DiVincenzo, and Andreas Winter for
helpful correspondence, comments, and discussions.

\appendix

\section{Properties of smooth cutoff functions}\label{app:cutoff}

In this Appendix we briefly review the properties of compactly
supported $C^\infty$ cutoff functions.

Of fundamental utility in our derivations is a class of functions
known as \emph{compactly supported $C^\infty$ bump functions}. These
functions are defined so that their fourier transform
$\widehat{\chi}_\gamma(\omega)$ is compactly supported on the
interval $[-\gamma, \gamma]$, and equal to $1$ on the middle third
of the interval. Such functions satisfy the following derivative
bounds
\begin{equation}\label{eq:chiderbound}
\frac{d^j\widehat{\chi}_\gamma(\omega)}{d\omega^j} \lesssim
\gamma^{-j},
\end{equation}
for all $j$ with the implicit constant depending on $j$. This is
just about the best estimate possible given Taylor's theorem with
remainder and the constraints that $\widehat{\chi}_\gamma(\omega)$
is equal to $1$ at $\omega = 0$ and $\widehat{\chi}_\gamma(\omega)$
is compactly supported.

The function $\chi_\gamma(t)$ has support throughout $\mathbb{R}$
but it is decaying rapidly. To see this consider
\begin{equation}
\chi_\gamma(t) = -\frac{1}{2\pi}\int_{-\infty}^{\infty} \frac{1}{it}
e^{-it\omega}\frac{d}{d\omega}\widehat{\chi}_\gamma(\omega) d\omega
\end{equation}
which comes from integrating by parts. Continuing is this fashion
allows us to arrive at
\begin{equation}
\chi_\gamma(t) = \frac{1}{2\pi}\int_{-\infty}^{\infty}
\left(-\frac{1}{it}\right)^j
e^{-it\omega}\frac{d^j}{d\omega^j}\widehat{\chi}_\gamma(\omega)
d\omega
\end{equation}
Since $\widehat{\chi}_\gamma(\omega)$ has all its derivatives
bounded, according to (\ref{eq:chiderbound}), and using the compact
support of $\widehat{\chi}_\gamma(\omega)$ we find
\begin{equation}
\begin{split}
|\chi_\gamma(t)| &\lesssim \left|\int_{-\gamma}^{\gamma}
\left(\frac{1}{it}\right)^j e^{-it\omega} \gamma^{-j} d\omega\right|
\\
&\lesssim \int_{0}^{\gamma} \frac{1}{|\gamma t|^j} d\omega \\
&\lesssim \frac{1}{\gamma^{j-1}|t|^j},
\end{split}
\end{equation}
for all $j\in \mathbb{N}$. Thus we find that $\chi_{\gamma}(t)$
decays to $0$ faster than the inverse of any polynomial in $t$ with
characteristic ``width'' $1/\gamma$. The existence and construction
of such functions is discussed, for example, in \cite{vaaler:1981a,
vaaler:1985a}.


\begin{thebibliography}{39}
\bibitem{endnote42}{At least those which are
equivalent to conformal models. Note that it is easy to construct
models which are ``critical'' in the sense of vanishing spectral
gap, but which require exponential resources to compute and store
their ground states.}
\bibitem{endnote44}{The class {\protect \sf  BQP} is
the class of decision problems which can be decided in polynomial
time on a quantum computer.}
\bibitem{endnote47}{That is, we assume
$|\protect \operatorname {supp}(A)|$ is any (arbitrarily large)
$O(1)$ constant.}
\bibitem{endnote48}{Note that this choice
precludes an extension of our analysis to study Berry phases in spin
systems. We'll relax this constraint in a future paper.}
\bibitem{endnote49}{Because of our $\lesssim $ notation we'll
reuse the symbol $c$ by systematically redefining it.}
\bibitem{endnote50}{Recall that the geometric entropy $S_\Lambda$ is
equal to the von Neumann entropy of the restriction of the ground
state $\rho(s) = |\Omega(s)\rangle \langle \Omega(s)|$ to a
contiguous block of $\Lambda$ spins: $S_\Lambda =
S(\rho_\Lambda(s))$, where $S(\rho) = -\tr(\rho \log_2(\rho))$ and
$\rho_\Lambda(s) = \tr_{\widehat{\Lambda}}(\rho(s))$, with
$\tr_{\widehat{\Lambda}}$ denoting partial trace over all spins
except those in $\Lambda$.}
\end{thebibliography}
\end{document}